\documentstyle[psfig]{mn}

\def\gs{\mathrel{\raise0.35ex\hbox{$\scriptstyle >$}\kern-0.6em
\lower0.40ex\hbox{{$\scriptstyle \sim$}}}}
\def\ls{\mathrel{\raise0.35ex\hbox{$\scriptstyle <$}\kern-0.6em
\lower0.40ex\hbox{{$\scriptstyle \sim$}}}}

\title[A test of popular models for symbiotic stars]
      {On the radio spectrum of CI~Cygni:
       a test of popular models for symbiotic stars}

\author[J.\ Miko{\l}ajewska \& R.\,J.\ Ivison]
       {J.\ Miko{\l}ajewska$^{\! 1}$\thanks{e-mail: mikolaj@camk.edu.pl}
        and R.\,J.\ Ivison$^{\! 2}$
        \vspace*{1mm}\\
        $^1$ N. Copernicus Astronomical Center, Bartycka 18, 00716 Warsaw, Poland\\
        $^2$ Department of Physics \& Astronomy, University College London,
             Gower Street, London WC1E 6BT}

\date{Accepted ... ;
      Received ... ;
      in original form ...}

\begin{document}

\maketitle

\begin{abstract}
We have determined the spectral energy distribution at wavelengths
between 6\,cm and 850\,$\mu$m for the proto-typical S(stellar)-type
symbiotic star, CI~Cygni, during quiescence. Data were obtained
simultaneously with the Very Large Array and the SCUBA sub-millimetre
(sub-mm) camera on the James Clerk Maxwell Telescope. The data have
allowed us to determine the free-free turnover frequency of the
ionised component, facilitating a model-dependent estimate of the
binary separation to compare with the known orbital parameters of
CI~Cyg and to critically test the known models for radio emission from
symbiotic stars. In particular, our data rule out the two most popular
models: ionization of the giant wind by Lyman continuum photons from
its hot companion and emission resulting from the interaction of winds
from the two binary components. \\

\end{abstract}


\section{Introduction}

The symbiotic stars are a collection of objects which exhibit the
characteristic optical--infrared absorption bands of a cool giant upon
which are superimposed emission lines whose origin requires conditions
typical of a much hotter object. Multi-frequency observations have led
to the universal acceptance that they are binary systems, typically
containing a post-asymptotic giant branch star or main-sequence dwarf
accreting matter from a late-type giant or supergiant.
 
Radio emission from symbiotic stars is dominated by free-free
radiation from gas ionised by the hotter of the binary
companions. Seaquist, Taylor \& Button (1984, hereafter STB) and
Seaquist et al.\ (1990, 1993) investigated the radio properties of
$\sim 100$ symbiotics, finding differences between the ionised regions
around D-type systems (those containing Miras, where the IR emission
is predominantly from dust) and S-type systems (containing
first-ascent giants where the IR emission is predominantly
photospheric). The SEDs of D-type systems turn over from partially
optically thick ($\alpha \sim 1$, where $F_{\nu} \propto
\nu^{\alpha}$) to optically thin emission ($\alpha \sim -0.1$) at
centimetre wavelengths, whereas the SEDs of S-type systems remain
optically thick throughout the radio regime.
 
The frequency of the turnover to optically thin free-free emission,
$\nu_{\rm t}$, is of particular interest: in the STB model, $\nu_{\rm
t} \propto a^{-0.7}$ (where $a$ is the binary separation) and the
optically thin portion of the free-free spectrum is very flat ($\alpha
\sim 0$); in the colliding winds model, $\nu_{\rm t} \propto
a^{-1.5}$, and the spectral index never flattens ($\alpha \sim +0.5$,
even for $\nu > \nu_{\rm t}$).

Mm/sub-mm observations have determined the spectral turnovers in
several D-type systems (Ivison, Hughes \& Bode 1992; Seaquist \&
Taylor 1992; Ivison et al.\ 1995). Until recently, however, the
limited sensitivity of bolometer detectors had hindered attempts to
pin down the turnover frequency in quiescent S-type systems. These are
thought to be fainter, probably because of a feedback between the
lower mass-loss rate of their earlier-type giant (compared with D-type
giants) and the consequent low accretion rate onto the ionising
companion star. The result is a lower gas density and a lower flux of
ionising radiation, which both lead to a lower emission
measure. Brighter S-type systems (e.g.\ AG~Peg, PU~Vul) are often
recovering from nova-type outbursts making them unsuitable
representives of the normally quiescent population.

%
%
\begin{table*}
\begin{center}
\caption{ \hfil Observed properties of CI~Cyg. \hfil }
\begin{tabular}{lcccl}
\noalign{\smallskip}
Property & Telescope & UT Date & Measurement & Comment \\
\noalign{\smallskip}
$\alpha$(J2000)& VLA & 1998 Feb 15 &$19^{\rm h} 50^{\rm m} 11.83^{\rm
s}$&From 3.54-cm image, $\pm0.1''$.\\
$\delta$(J2000)&     &             &$+35^{\circ} 41' 03.2''$&East of
position reported by Seaquist et al.\ (1993).\\
\noalign{\smallskip}
3600\,\AA\ ($U$) & CAO  & 1985 Jun 07      & $16.8\pm0.1$\,mJy  & Munari
et al.\ (1992).\\
4400\,\AA\ ($B$) &      &                  & $44.4\pm0.1$\,mJy  &\\
5500\,\AA\ ($V$) &      &                  & $0.16\pm0.01$\,Jy  &\\
7000\,\AA\ ($R$) &      &                  & $0.77\pm0.01$\,Jy  &\\
1.25\,$\mu$m ($J$)&      &                  & $8.02\pm0.07$\,Jy &\\
1.65\,$\mu$m ($H$)&      &                  & $10.4\pm0.1$\,Jy  &\\
2.20\,$\mu$m ($K$)&      &                  & $10.1\pm0.1$\,Jy  &\\
\noalign{\smallskip}
12\,$\mu$m   &{\em IRAS}& 1984         & $0.93\pm0.06$\,Jy
&Whitelock \& Munari (1992).\\
25\,$\mu$m   &          &              & $0.33\pm0.03$\,Jy&\\
60\,$\mu$m   &          &              & $3\sigma<0.45$\,Jy&\\
100\,$\mu$m  &          &              & $3\sigma<5.97$\,Jy&\\
\noalign{\smallskip}
450\,$\mu$m  & JCMT & 1998 Jan 29, 31  & $3\sigma<42.7$\,mJy
&Photometry mode.\\
850\,$\mu$m  &      &                  & $9.50\pm1.29$\,mJy&\\
1350\,$\mu$m &      & 1998 Jan 30      & $10.1\pm1.4$\,mJy&\\
2000\,$\mu$m &      &                  & $11.0\pm2.8$\,mJy&\\
\noalign{\smallskip}
0.69\,cm     & VLA  & 1998 Feb 20      & $7.35\pm1.55$\,mJy  &From raw
0-150\,k$\lambda$ visibilities.\\
1.33\,cm     &      &                  & $3.92\pm0.75$\,mJy  &\\
2.01\,cm     &      & 1998 Feb 15      & $3.01\pm0.29$\,mJy  &\\
3.54\,cm     &      &                  & $1.49\pm0.10$\,mJy  &From map
made with 0-150\,k$\lambda$ visibilities.\\
6.17\,cm     &      &                  & $0.91\pm0.08$\,mJy  &\\
\end{tabular}
\end{center}
\end{table*}

This situation is unfortunate because the STB model can be tested
against the SEDs of quiescent systems: a critical parameter in the
model is the binary separation, which is known for many S-type
symbiotic binaries (where periods are a few years, typically) but not
known for D-type systems (which are thought to be more widely
separated and to have periods spanning one or more decades).

Attempts {\em have} been made to constrain $\nu_{\rm t}$ for several
well-studied S-type systems (RW~Hya --- Miko{\l}ajewska \& Omont 1998;
AG~Peg and Z~And --- Ivison et al.\ 1992, 1995) but these studies have
usually had to rely on estimates of the optically thin free-free
emission in the sub-mm range derived from optical/ultraviolet (UV)
H\,{\sc i} free-free and bound-free emission measures. The consistency
between the data and binary models revealed by the aforementioned work
is not therefore based on direct measurements of the turnover
frequency, only on an optically thin emission measure inferred from
optical/UV spectroscopy. We need to sample the mm/sub-mm spectral
energy distribution of S-type symbiotic stars in order to define the
turnover frequency in a model-independent manner.

For the investigation reported here, we selected the well-studied
S-type system, CI~Cyg, a proto-typical symbiotic star with a known
orbital period ($P_{\rm orb} = 855.6$\,d) and a well-defined
spectroscopic orbit (Miko{\l}ajewska 1997; Belczy{\'n}ski et al.\ 2000
and references therein). A thorough spectroscopic study of CI~Cyg by
Kenyon et al.\ (1991) demonstrated that it consists of an M5\,II
asymptotic branch giant, $M_{\rm g} \sim 1.5\, \rm M_{\sun}$, and a
$\sim 0.5\, \rm M_{\sun}$ companion separated by 2.2\,{\sc au}.  Our
objective was to define the shape of the spectral energy distribution
from centimetre (cm) to sub-mm wavelengths, determining $\nu_{\rm t}$
and thus quantitatively testing and discriminating between the
existing models for symbiotic binaries. In \S2, we report our
measurements of CI~Cyg, which we obtained near-simultaneously at
wavelengths between 850\,$\mu$m and 6\,cm during 1998 January and
February. Our interpretation of the resulting spectral energy
distribution is presented in \S3.

\section{Observations}

\subsection{Measurements at 0.7---6.2\,cm with the VLA}

Observations were carried out during 1998 February 15 and 20 ({\sc
ut}) with the NRAO Very Large Array (VLA), New Mexico, during a move
between the D and A configurations.  Around 30\,min was spent on
source at 0.69, 1.33, 2.01, 3.54 and 6.17\,cm, with measurements of
the bright gain/phase calibrator, 2015+371, every 5--10\,min. The flux
density scale was set using observations of 3C\,286 at 2.01, 3.54 and
6.17\,cm, and using a variety of bright calibrators at 0.69 and
1.33\,cm. We recorded a bandwidth of 100\,MHz -- two contiguous 50-MHz
bands, with both right and left circular polarizations.

Calibration of the synthesis data followed standard NRAO Cookbook
procedures within {\sc aips}. At 6.17 and 3.54\,cm, the {\sc mx}
routine was used to make and {\sc clean} maps, restricting the $uv$
coverage to 0--150\,k$\lambda$ to reject those antennas which had
already been moved into A-configuration positions. At shorter
wavelengths we utilised prior knowledge of the source position (from a
3.54-cm map made using all baselines) to determine the flux density
directly from the visibility data, again restricting ourselves to
short baselines typical of D configuration. The measured flux
densities are listed in Table~1.

\subsection{Measurements at 450---2000\,$\mu$m with SCUBA on the JCMT}

Observations with SCUBA (Holland et al.\ 1999) on the 15-m James Clerk
Maxwell Telescope (JCMT), Mauna Kea, Hawaii, were made using its
photometry mode at 450, 850, 1350 and 2000\,$\mu$m during the nights
of 1998 January 29--31 ({\sc ut}). The secondary mirror was chopped in
azimuth by 60$''$ at 7.8\,Hz and jiggled every 1\,s in a simple
9-point pattern with 2$''$ offsets. The telescope was nodded between
the signal and reference beams every 9\,s in a
signal--reference--reference--signal pattern. The pointing was checked
regularly and skydips were performed every hour to measure the
atmospheric opacity.

Reducing the data, spikes were carefully rejected then measurements in
the reference beam were subtracted from those in the signal beam. The
data were corrected for atmospheric opacity and calibrated against
Uranus and Mars. The measured flux densities are listed in Table~1.

\section{Results and discussion}

The radio spectrum of CI~Cyg is shown in Table~1 and Fig.~1 together
with earlier radio data (Torbett \& Campbell 1989; Seaquist, Krogulec
\& Taylor 1993) and published optical/IR measurements (Munari et al.\
1992; Whitelock \& Munari 1992). The resulting spectral energy
distribution covers the entire radio to optical range. The IR region
is clearly dominated by photospheric emission (from an M5 III giant --
solid line in Fig.~1) while the cm to sub-mm emission is dominated by
partially optically thick bremsstrahlung (free-free) from ionised gas.
The radio spectrum turns over at mm wavelengths indicating that it has
become optically thin to free-free emission.

To estimate the turnover frequency, $\nu_{\rm t}$, the (partially)
optically thick cm-wave radio data were fitted with a power law
spectrum, $S_{\nu} \propto \nu^{\alpha}$. The optically thin sub-mm
data were similarly fitted but the spectral index was arbitrary fixed
at $\alpha = -0.1$ (appropriate for optically thin emission). For
$\nu_{\rm t}$, we have adopted the point at which the low-frequency
extrapolation of the optically thin free-free emission exceeds the
observed optically thick emission by a factor $e$ ($\tau_{\nu}=1$). A
least-squares fit to our data yields $S_{\nu}^{\rm thick} =
0.20\pm0.02 (\nu/\rm GHz)^{0.96(\pm0.03)}$\,mJy and $S_{\nu}^{\rm
thin} = 17.5 \pm0.6 (\nu/\rm GHz)^{-0.1}$\,mJy for the cm and sub-mm
data, respectively, and we estimate $\nu_{\rm t} = 27 \pm 5$\,GHz.

%
%
\begin{figure}
\centerline{\psfig{file=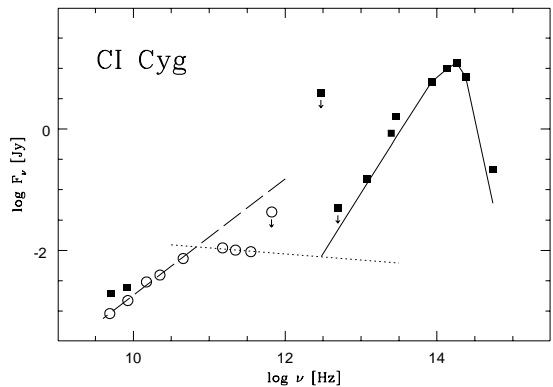,width=7.2cm}}
\caption{The spectral energy distribution of CI~Cyg between the radio
and optical wavebands. New measurements are represented by empty
circles, old measurements by filled squares. The dashed and the dotted
lines, respectively, represent power laws with spectral indices
$+0.96$ and $-0.10$ (see text).}
\end{figure}

For free-free emission in the optically thick regime, the effective
radio photosphere emitting at frequency $\nu$ and temperature $T$ is
given approximately by the blackbody radius,
\begin{equation}
\frac{R}{\sc au} = 210 \left(\frac{S_{\nu}}{\rm mJy}\right)^{0.5} 
\left(\frac{T}{10^4\,{\sc k}}\right)^{-0.5} 
\left(\frac{\nu}{\rm GHz}\right)^{-1} 
\left(\frac{d}{\rm kpc}\right).
\end{equation}
In the case of CI~Cyg, the radius at the optically thick-thin
transition (i.e.\ $\nu = \nu_{\rm t}$) is $52.6\,(T/10^4\,{\sc
k})^{-0.5}$ {\sc au}, adopting a distance of 2\,kpc (Kenyon et al.\
1991), which should represent the inner radius of its optically thick
shell.  This radius is significantly larger than the size of the known
binary orbit of CI~Cyg, for any reasonable temperature.  The emission
measure of the shell, derived from the optically thin turnover
frequency, $\nu_{\rm t} \sim 27$\,GHz, is around $6 \times
10^{14}\,(T_{\rm e}/10^4{\sc k})^{1.35}$\,cm$^{-6}$\,{\sc au} measured
along the line-of-sight through its densest region.

Finally, the optically thin flux density is consistent with a volume
emission measure of $n_{\rm e}^2\,V \approx 3$ to $5 \times 10^{58}
(d/\rm 2\,kpc)^2\,\rm cm^{-6}$ for the respective range of $T_{\rm e}
\sim 10^{4}$ to $10^{5}$\,{\sc k} found in CI~Cyg (Miko{\l}ajewska
1985; Kenyon et al.\ 1991). This emission measure is an order of
magnitude lower than the emission measure of $3$ to $10 \times
10^{59}\,\rm cm^{-6}$ derived from the optical/UV bound-free and
free-free continuum and Balmer H\,{\sc i} emission lines, adopting a
reasonable range for $E(B-V)$ ($\sim$0.3--0.45), thus the free-free
radio emission and the hydrogen optical/UV recombination emission from
CI~Cyg do not seem to arise from the same volume of gas.

The radio emission also provides a model-independent lower limit to
the Lyman continuum photon luminosity, $L_{\rm Ly}$:
\begin{equation}
\geq 7 \times 10^{43}\,\frac{S_{\nu}}{\rm mJy}\,\left(\frac{T_{\rm e}}{10^4\,{\sc k}}\right)^{-0.4}\,\left(\frac{\nu}{\rm GHz}\right)^{0.1}\,\left(\frac{d}{\rm kpc}\right)^2 \rm s^{-1}
\end{equation}
(Seaquist \& Taylor 1987), which gives $L_{\rm Ly} \gs 5$ to $2 \times
10^{45}$\,phot\,s$^{-1}$ for the appropriate range of $T_{\rm e}$
($10^4$ to $10^5$), much lower than the $L_{\rm Ly} \sim
10^{47}$\,phot\,s$^{-1}$ indicated by studies of the UV continuum and
He\,{\sc ii} emission lines observed during quiescence (1985-94;
Miko{\l}ajewska \& Ho{\l}owacz 2000).

The current radio spectrum may be used to test the known models for
radio emission from symbiotic stars.  Above all, the very flat shape
of the optically thin portion of our spectrum rules out the colliding
wind model (in which the spectral index is never expected to flatten:
$\alpha \sim +0.5$, even for $\nu > \nu_{\rm t}$ (Kenny 1995).

\subsection{Single-star models}

The very steep optically thick spectral index is also inconsistent
with an isothermal, spherically symmetric steady flow (for which
$\alpha = +0.6$, Wright \& Barlow 1975).  There are several possible
reasons for such a steep spectral index.  For example, Seaquist \&
Taylor (1987) showed that in the case of a wind with a power-law
temperature and density profile of the form $T \propto r^{\rm -n}$ and
$n \propto r^{\rm -p}$, respectively, the observed spectral index is
given by
\begin{equation}
\alpha = \frac{4p - 6.2 - 0.6 n}{2p - 1 - 1.35 n}.
\end{equation}
In principle, then, $\alpha = +0.96$ can be accounted for by a steady
wind ($p=2$) and a temperature gradient with a power-law index,
$n=1.55$, or by an isothermal wind with a density gradient, $p=2.51$.

Seaquist \& Taylor (1987) noted that the optically thick portion of
the radio spectrum would be steeper than that expected for a steady,
spherically symmetric flow if the object underwent episodic mass loss
(i.e.\ if the flow was not steady). The turnover of the spectrum at
high frequencies is then due to the finite inner boundary of a
shell-like emission region caused by the episodic mass loss.

Although only a few attempts have been made to detect CI~Cyg at radio
wavelengths, it seems that the radio emission has been variable over
at least last two decades (Table~2) and this variability could in
principle be related to the large eruption that gave rise to an
optical maximum in 1975. CI~Cyg was not detected at 6\,cm in 1982
February, whereas in 1987 October its 6-cm flux density was a factor
of $4 - 5$ stronger than the $3\sigma$ upper limit from 1982 (Table~2)
and a factor $\sim 2$ higher than our present measurement. It seems
unlikely, however, that the radio-emitting gas was ejected in 1975
because the delay between the maximum of the optical/UV light curve
and the onset of the radio emission is too long. For comparison, this
delay was of order of days (UV) to months (optical) for the outburst
of Z~And (Fernandez-Castro et al.\ 1995).  Also, the present
characteristic radius at the optically thin-thick transition cannot
represent the inner radius of an optically thick shell ejected over 20
years ago unless the expansion velocity of this shell is extremely low
($\ls 13\,\rm km\,s^{-1}$).
Such a low expansion velocity is, of course, compatible 
with a wind from the cool giant and a radio spectrum
of  the form observed for CI Cyg can be produced by a discontinuity
in that wind.
However, we consider such a scenario to be unlikely.
In particular, if the 1975 outburst
were due to an episodic increase in the cool giant wind (as in
the model proposed by Nussbaumer \& Vogel 1987)
changes in the optical and near-IR magnitudes of the giant should
have been observed, contrary to the observations 
(Miko{\l}ajewska \& Kenyon 1992).

The respective absence and presence of radio emission in 1982 and 1987
may be related to the spectral changes in the optical/UV continuum and
emission-line light curves observed in the 1980s (Miko{\l}ajewska
1985; Kenyon et al.\ 1991).  In particular, in 1975-83 (during the
outburst and its decline) eclipses in $UBV$ continuum and optical
H\,{\sc i}, He\,{\sc i} and He\,{\sc ii} emission lines were narrow,
with a total duration of $\ls 0.16\,P_{\rm orb}$ with well-defined
eclipse contacts. Miko{\l}ajewska (1985) noted that the radii of the
H$^+$ and He$^{+2}$ regions in the orbital plane (as derived from the
eclipse profiles) were a factor of $2-3$ smaller than the respective
spherical radii derived from the observed emission measures,
indicating that the emission may arise in a non-spherical (perhaps
bipolar) nebula or flow.  Since 1984, the eclipses in both the
continuum and the emission lines (except for He\,{\sc ii}, where the
eclipse behaviour has remained practically unchanged) have become very
broad and the light curves show continuous sinusoidal variations
suggesting that the bulk of the Balmer H\,{\sc i} and He\,{\sc i}
emission is coming from a region between the binary components,
possibly with a density concentration towards the cool giant and in
the orbital plane.

The transition from narrow eclipses to sinusoidal variations was
accompanied by an increase by a factor of $\sim 2$ in Balmer H\,{\sc
i} line and continuum emission, a decline in [O\,{\sc iii}] and
[Ne\,{\sc iii}] emission and the appearance of very strong, highly
ionised [Fe\,{\sc vii}], [Ne\,{\sc v}] and [Mg\,{\sc v}] emission
lines in the optical and UV bands.  At the same time, both the eclipse
behaviour and the intensity of He\,{\sc ii} lines remained unchanged.
The physical conditions in the nebular region also changed: the
intensity ratios of the [O\,{\sc iii}] and [Ne\,{\sc iii}] lines were
consistent with a low-density photoionised region, $n_{\rm e} \sim
{\rm a\,\,few}\, \times 10^6\, \rm cm^{-3}$, $T_{\rm e} \sim 10\,000 -
20\,000\, \sc k$, while the high-ionization emission lines required a
low-density, hot, $n_{\rm e} \ls 10^7\, \rm cm^{-3}$, $T_{\rm e} \sim
10^5$\,{\sc k}, mechanically-heated region.  Kenyon et al.\ (1991),
Miko{\l}ajewska \& Kenyon (1992) associated these changes with a
transition from an optically thick to an optically thin accretion disc
following the decline from the 1975 eruption and suggested that the
high-ionization forbidden line emission might be excited in an
accretion disk corona that formed when the outer disk became optically
thin.  This transition occured within a year or so, and the [Fe\,{\sc
vii}] and [Mg\,{\sc v}] emission lines remained strong until at least
1994 (Miko{\l}ajewska \& Ho{\l}owacz 2000).  Kenyon et al.\ (1991)
estimated the spherical radius of this highly ionised region as $R
\sim 25\, (d/2\,\rm kpc)^{2/3}\, \sc au$, which is surprisingly close
to the blackbody radius, $R \sim 37\,(T_{\rm
e}/10^5)^{-0.5}\,(d/2\,\rm kpc)\, \sc au$ appropriate for the 4.9-GHz
flux density reported by Torbett \& Campbell (1989).  If these
connections between the radio emission and the spectral changes are
not coincidental then the high-ionization [Fe\,{\sc vii}], [Ne\,{\sc
v}] and Mg\,[{\sc v}] lines and the radio emission might have a common
origin.

\begin{table}
\begin{center}
\caption{History of radio observations of CI~Cyg}
\label{}
\begin{tabular}{@{}lccc@{}}
\hline
Date & Freqency & Flux & Reference\\
     & (GHz)  & (mJy) & \\
\hline
1982 Feb 9 & ~4.9 & $<0.42$ & ST90\\
1987 Oct   & ~4.9 & $1.92\pm0.04$ & TC89\\
1988 Oct 14& ~4.9 & $<1.08$ & I\&91 \\
1989 Oct 4 & ~8.3 & $2.47\pm0.15$ & S\&93\\
1994 Apr 7 & 88.2 & $<7.1$ & I\&95\\
1998 Feb 15&  4.9 & $0.91\pm0.08$ & Here\\
\hline
\end{tabular}
\end{center}
\smallskip
{\footnotesize 
ST90 -- Seaquist \& Taylor 1990; TC89 -- Torbett \& Campbell 1989;
I\&91 -- Ivison et al.\ 1991; S\&93 -- Seaquist et al.\ 1993;
I\&95 -- Ivison et al.\ 1995.}
\end{table}

\subsection{Binary-star models}

Another possible scenario involves an asymmetric, ionised region
produced when a stellar wind is ionised by a hot companion, as in the
STB model. The geometry of the ionised region is governed by a single
parameter, $X$ (the ionization parameter), which depends on the red
giant mass-loss rate, the binary separation, and the Lyman continuum
luminosity of the ionizing source.  In this model, knowledge of the
turnover frequency, $\nu_{\rm t}$, can be used to estimate the binary
separation, $a$, which can then be compared with independently-known
values (Table 3).  For a range in $X$ covering two orders of magnitude, the
binary separation (within a factor of 2) is (Seaquist \& Taylor 1990):
\begin{equation}
\frac{a}{\sc au} = 300 \left(\frac{T_{\rm e}}{10^4\,\sc k}\right)^{-0.5
}\,\left(\frac{\nu_{\rm t}}{\rm GHz}\right)^{-1}\,\left(\frac{S_{\rm t}}{\rm mJy}\right)^{0.5} \left(\frac{d}{\rm kpc}\right),
\end{equation}
where $S_{\rm t}$ is the optically thin flux near the turnover, and
$d$ is the distance.  For $\nu_{\rm t} = 27$\,GHz and $S_{\rm t} \sim
12.6$\,mJy, we find $a_{\rm radio} \sim 40 (d/\rm kpc)$\,{\sc au}.  A
comparison with the binary separation derived from spectroscopic orbit
solution, $a_{\rm sp} \sim 2.2$\,{\sc au} (Kenyon et al.\ 1991) shows
that the observed turnover frequency over-estimates the binary
separation by a factor of $\sim 30$ for $d \sim 2$\,kpc. The STB model
apparently does not account satisfactorily for the radio emission from
CI~Cyg.

\begin{table}
\begin{center}
\caption{Binary parameters of CI~Cyg}
\label{}
\begin{tabular}{@{}lcc@{}}
\hline
& Adopted value & Reference\\
\hline
Distance [pc] & 2000 & K\&91\\
Orbital period [days] & 855.6 & B\&00\\
Mass ratio, $M_{\rm g}/M_{\rm h}$ & 3 & K\&91\\
Inclination & $\ga 73^{\circ}$ & K\&91\\
Separation of the stars [AU] & 2.2 & K\&91\\
Mass of the red giant [M$_{\sun}$] & 1.5 & K\&91\\
Mass of the hot component [M$_{\sun}$] & 0.5 & K\&91\\
\hline
\end{tabular}
\end{center}
\smallskip
{\footnotesize 
K\&91 -- Kenyon et al.\ 1991; B\&00 -- Belczy{\'n}ski et al.\ 2000.}
\end{table}

There are also other inconsistencies between the observed spectrum of
CI~Cyg and the STB model.  We have already mentioned (see \S3 above)
the serious disagreement between the Lyman continuum luminosity
derived from the optically thin radio emission and that estimated from
the UV continuum and emission-line studies.  In addition, the observed
optically thick spectral index ($\alpha>+0.6$) immediately implies $X
< \pi/4$ in the STB model.  Since CI~Cyg is a well-studied eclipsing
system we know that our radio observations were made when the system
was viewed along the binary axis with the hot star in front (i.e.\ the
viewing angle, as defined in STB, $\theta \approx 0^{\circ}$).  Thus
$\alpha = +0.96$ indicates $X \approx 0.7$ (c.f.\ Fig.~3 of Seaquist
\& Taylor 1984).  The nebula in CI~Cyg should thus be radiation
bounded. This is inconsistent with its observed forbidden-line
spectrum. In particular, after 1984 the only strong lines have been
those with high excitation ([Fe\,{\sc vii}], [Ne\,{\sc v}] and
[Mg\,{\sc v}]) while the intermediate- and low-excitation lines have
been very faint ([O\,{\sc iii}], [Ne\,{\sc iii}]) or absent (e.g.\
[O\,{\sc ii}], [S\,{\sc ii}]) suggesting that the nebula is density
bounded ($X > 1$).  Further insight into the geometry of the nebular
region(s) of CI~Cyg is provided by the apparent lack of
Raman-scattered O\,{\sc vi} lines at $\lambda\lambda 6825, 7082$.
These lines, commonly observed in high-excitation symbiotic systems,
have {\it never} been detected in CI~Cyg, even at the epoch when the
high-ionization lines were very strong. A plausible explanation seems
to be a lack of sufficient neutral H$^0$ scatterers, which also
implies a high value of $X$, ($\ga 10$).

There are several plausible explanations for the failure of the STB
model in the case of CI~Cyg.  Above all, CI~Cyg is one of the few
symbiotic binaries in which the cool giant fills or nearly fills its
Roche lobe (e.g. Miko{\l}ajewska 1996, and references therein).  This
has two important implications: first, the size of the giant fills a
large fraction of the binary separation ($\sim 0.47a$ for $q \equiv
M_{\rm g}/M_{\rm h} \sim 3$) and for $X \la 1/3$ the ionisation front
impinges on its photosphere; second, and perhaps most important, mass
loss from the giant is strongly concentrated in a stream flowing
through the inner langrangian point ($L_1$) towards the hot component
(as opposed to the spherically symmetric wind assumed in the STB
model).  The result is a complex density distribution, not the simple
$\sim r^{-2}$ distribution in the STB model.  In particular, the
density should vary roughly as $r^{-1}$ along the stream, possibly in
addition to a more symmetric component with much lower density -- an
order of magnitude or more lower that that in the stream (e.g.\ Lubov
\& Shu 1975).  Thus the simple geometry assumed for the STB model is
incompatible with the conditions in the circumbinary environment of
CI~Cyg.

The fact that the giant component of CI~Cyg fills its tidal lobe also
implies that the interacting wind model (assuming spherically
symmetric winds from both components) cannot be a good match to the
true conditions -- the stream should result in the formation of a disk
of material orbiting the companion. Any wind from the companion will
thus be bipolar rather than spherically symmetric.

Radio continuum spectra of collimated ionised stellar winds have been
calculated by Reynolds (1986) showing that unresolved sources can have
partially opaque spectra with spectral indices between $+2$ and
$-0.1$. These models assume the thermal material appears at a finite
radius, $r_{\rm 0}$, where the jet half-width is $w_{\rm
0}$. Power-law dependencies of quantities on jet length, $r$, are also
assumed; in particular, the jet half-width, temperature, velocity,
density and ionised fraction are assumed to vary as $r/r_{\rm 0}$ to
powers of $\epsilon$, $q_{\rm T}$, $q_{\rm v}$, $q_{\rm n}$, and
$q_{\rm x}$, respectively. The models are defined by values of
$\epsilon$, $q_{\rm v}$, and $q_{\rm x}$; the other gradients are
derived from these.  In particular, a spectrum with $\alpha = +0.95$,
as observed for CI~Cyg, can be produced by some of the models. A good
example is a conical, recombining, cooling, constant-velocity jet: the
observed spectral index can be reproduced by the following combination
of gradients: $q_{\rm v}=0$, $\epsilon =1$, $q_{\rm x}=q_{\rm T}=-1/3$
(using Reynolds' notation). Then $q_{\rm \tau}=-3.22$, and $F(q_{\rm
\tau}, \alpha_{\rm op})=1.24$. 
Assuming the jet is perpendicular to
the orbital plane of CI~Cyg (and the eclipses observed in CI~Cyg do
imply $i \sim 90^{\circ}$) and using 
our observed values of the turnover frequency and the radio flux 
in equations (18) and (19) from
Reynolds, we obtain a characteristic jet extension (collimation 
distance) and mass-loss rate
of
\begin{equation}
r_0 \sim 133 \left(\frac{\Theta_0}{0.4}\right)^{-0.5}
\left(\frac{T_{\rm e}}{10^4}\right)^{-0.5}\,{\sc au},
\end{equation}
and
\begin{equation}
\dot{M} \sim 2.85 \times 10^{-7} \left(\frac{v}{200\,\rm km\,s^{-1}}\right)
\left(\frac{\Theta_0}{0.4}\right)^{0.75}
\rm M_{\odot}\,yr^{-1}.
\end{equation}
The jet would thus have an extension of $2\,r_0/d \sim 133/42$
milliarcsec for $T \sim 10^4/10^5$\,{\sc k}, respectively, and in
principle should be detectable.

It is interesting that for a reasonable $\Theta$ ($\sim 0.4$,
consistent with the ratio of the equatorial to spherical radius of
$H^+$ and He$^{+2}$ estimated from narrow eclipse features -- see
\S3.1), the jet cuts out $(\Theta/2)^2 \sim 0.04$ of the solid angle,
comparable to the ratio of the Lyman continuum luminosities as
derived from the radio and UV data, respectively.

According to Reynolds' models, the spectral index of the integrated
flux reflects the gradients in the region of the flow 
corresponding to the frequency of observation; 
spatial structure in the source is thus reflected in the
spectral signature of the flux. 
In general, the conditions may vary along the flow, resulting
in steepening and flattening trends in spectra -- 
an example is shown in Figure 2 of Reynolds (1986).
If the inner jet of CI Cyg is confined ($\epsilon < 1$)
it can be responsible for the flat spectrum (optically thin) 
between 2 and 0.85\,mm, whilst the cm range would be dominated
by the free ($\epsilon = 1$) outer jet.

\section{Conclusions}

We conclude that neither the STB nor the colliding winds model can
account satisfactorily for the radio emission from CI~Cyg. This is
probably because the cool component in this binary fills its Roche
lobe so that the geometry of the mass flow differs significantly from
the simple geometries assumed for those models.

It seems that the bulk of the Balmer H\,{\sc i} emission is formed in
a dense region between the binary components, probably in the
partially ionised mass-loss stream, while the high-excitation
forbidden lines and radio emission arises from lower density regions,
perhaps in a bipolar outflow.

\subsection*{ACKNOWLEDGEMENTS}

RJI acknowledges the award of a PPARC Advanced Fellowship. This study
was supported in part by the KBN Research Grant No 2P03D\,021\,12.


\begin{thebibliography}{99}

\bibitem{} Belczy{\'n}ski, K., Miko{\l}ajewska, J., Munari, U.,
           Ivison, R.J., Friedjung, M., 2000, A\&AS, 146, 407{}{}
\bibitem{} Fernandez-Castro T., Gonzalez-Riestra R., Cassatella A., 
           Taylor A.R., Seaquist R.E., 1995, ApJ, 442, 366{}{}
\bibitem{} Holland W.S., Robson E.I., Gear W.K.\ et al., 1999, MNRAS,
           303, 659{}{}
\bibitem{} Ivison R.J., Bode M.F., Roberts J.A., Meaburn J.,
           Davis R.J., Nelson R.F., Spencer R., 1991, MNRAS, 249, 347{}{}
\bibitem{} Ivison R.J., Hughes D.H., Bode M.F., 1992, MNRAS, 257, 47{}{}
\bibitem{} Ivison R.J., Seaquist E.R., Schwarz H.E., Hughes D.H., Bode
           M.F., 1995, MNRAS, 273, 517{}{}
\bibitem{} Kenny H.T., 1995, PhD Thesis, Univ. of Calgary
\bibitem{} Kenyon S.J., Oliversen N.A., Miko{\l}ajewska J., 
           Miko{\l}ajewski M., Stencel R.E., Garcia M.R., Anderson C.M.,
           1991, AJ, 101, 637{}{}
\bibitem{} Lubov S.H., Shu F., 1975, ApJ, 198, 383{}{}
\bibitem{} Miko{\l}ajewska J., 1985, Acta Astr., 35, 65{}{}
\bibitem{} Miko{\l}ajewska J., 1996, in Evans A., Wood J.H., eds, 
           Cataclysmic Variables and Related Objects, Kluwer, Dordrecht, p.335
\bibitem{} Miko{\l}ajewska J., 1997, in Miko{\l}ajewska J., ed,
           Physical Processes in Symbiotic Binaries and Related Systems.
           Copernicus Foundation for Polish Astronomy, Warsaw, p.3{}{}
\bibitem{} Miko{\l}ajewska J., Ho{\l}owacz S., 2000, in preparation 
\bibitem{} Miko{\l}ajewska J., Kenyon S.J., 1992, MNRAS, 256, 177
\bibitem{} Miko{\l}ajewska J., Omont A., 1998, in K.L. Chan et al, eds,
           1997 Pacific Rim Conference on Stellar Astrophysics, 
           ASP Conf. Ser. Vol.138, p.241{}{}
\bibitem{} Munari U., Yudin B.F., Taranova O.G.\ et al., 1992,
           A\&AS, 93, 383{}{}
\bibitem{} Nussbaumer, H., Vogel, M., 1987, A\&A, 182, 51
\bibitem{} Reynolds S.P., 1986, ApJ, 304, 713{}{}
\bibitem{} Seaquist E.R., Krogulec M., Taylor A.R., 1993, ApJ, 410, 260{}{}
\bibitem{} Seaquist E.R., Taylor A.R., 1987, ApJ, 312, 813{}{}
\bibitem{} Seaquist E.R., Taylor A.R., 1990, ApJ, 349, 313{}{}
\bibitem{} Seaquist E.R., Taylor A.R., 1992, ApJ, 387, 624{}{}
\bibitem{} Seaquist E.R., Taylor A.R., Button S., 1984, ApJ, 284, 202 (STB){}{}
\bibitem{} Torbett M.R., Campbell B., 1989, ApJL, 340, L73{}{}
\bibitem{} Whitelock P.A., Munari U., 1992, A\&A, 255, 171{}{}
\bibitem{} Wright A.E., Barlow M.J., 1975, MNRAS, 170, 41{}{}

\end{thebibliography}
\end{document}